\documentclass[aps,prd,reprint,showpacs,showkeys,superscriptaddress,preprintnumbers,nofootinbib]{revtex4-1}
\usepackage{amsmath,amssymb,bm,mathrsfs}
\usepackage[dvips]{hyperref}
\usepackage{url}
\usepackage{breakurl}
\usepackage{graphicx}

\begin{document}

\preprint{UCB-PTH-12/08}
\preprint{IPMU12-0127}

\title{Compact Supersymmetry}

\author{Hitoshi Murayama}
\affiliation{Berkeley Center for Theoretical Physics, Department of Physics, 
  University of California, Berkeley, CA 94720, USA}
\affiliation{Theoretical Physics Group, Lawrence Berkeley National 
  Laboratory, Berkeley, CA 94720, USA}
\affiliation{Kavli Institute for the Physics and Mathematics of the 
  Universe (WPI), Todai Institutes for Advanced Study, University of Tokyo, 
  Kashiwa 277-8583, Japan}

\author{Yasunori Nomura}
\affiliation{Berkeley Center for Theoretical Physics, Department of Physics, 
  University of California, Berkeley, CA 94720, USA}
\affiliation{Theoretical Physics Group, Lawrence Berkeley National 
  Laboratory, Berkeley, CA 94720, USA}

\author{Satoshi Shirai}
\affiliation{Berkeley Center for Theoretical Physics, Department of Physics, 
  University of California, Berkeley, CA 94720, USA}
\affiliation{Theoretical Physics Group, Lawrence Berkeley National 
  Laboratory, Berkeley, CA 94720, USA}

\author{Kohsaku Tobioka}
\affiliation{Kavli Institute for the Physics and Mathematics of the 
  Universe (WPI), Todai Institutes for Advanced Study, University of Tokyo, 
  Kashiwa 277-8583, Japan}
\affiliation{Department of Physics, University of Tokyo, Hongo, 
  Tokyo 113-0033, Japan}

\begin{abstract}
 Supersymmetry broken geometrically in extra dimensions naturally leads 
 to a nearly degenerate spectrum for superparticles, ameliorating the 
 bounds from the current searches at the LHC.  We present a minimal 
 such model with a single extra dimension, and show that it leads to 
 viable phenomenology despite the fact that it essentially has two less 
 free parameters than the conventional CMSSM.  The theory does not suffer 
 from the supersymmetric flavor or $CP$ problem because of universality 
 of geometric breaking, and automatically yields near-maximal mixing 
 in the scalar top sector with $|A_t| \approx 2 m_{\tilde{t}}$ to boost 
 the Higgs boson mass.  Despite the rather constrained structure, the 
 theory is less fine-tuned than many supersymmetric models.
\end{abstract}

\maketitle

{\it Introduction.}~---~Supersymmetry has widely been regarded as the 
prime candidate for physics beyond the standard model~\cite{Martin:1997ns}. 
It can explain the dynamical origin of electroweak symmetry breaking 
through renormalization group effects and provide a natural candidate 
for the cosmological dark matter in its simplest incarnations.  In 
particular, it stabilizes the large hierarchy between the electroweak 
scale $\approx~{\rm TeV}$ and the quantum gravity scale $\approx 
10^{15}~{\rm TeV}$ against radiative corrections to the Higgs mass 
parameter.  Barring a fine-tuning among parameters of the theory, 
this consideration strongly suggests the existence of superparticles 
below $\approx ~{\rm TeV}$.  The mass spectrum of superparticles 
has been mostly discussed within the ``Minimal Supergravity'' 
or Constrained Minimal Supersymmetric Standard Model (CMSSM) 
framework~\cite{Chamseddine:1982jx}, which typically generates 
a widely spread spectrum leading to experimentally identifiable 
large visible and missing energies.

However, no experimental hints have been seen so far at the Large Hadron 
Collider (LHC), which has led to substantial anxiety in the community. 
Moreover, the suggested mass of $125~{\rm GeV}$ for the Higgs boson by 
the LHC data~\cite{ATLAS:2012ae} is not easily accommodated in the Minimal 
Supersymmetric Standard Model (MSSM), where one has to rely on radiative 
corrections to push the Higgs boson mass beyond the tree-level upper 
bound of $m_Z \simeq 91~{\rm GeV}$.  These requirements push the scalar 
quark masses well beyond the TeV within the CMSSM.

There are three main suggestions to allow for supersymmetry without 
the signal so far, within the context of $R$-parity conserving 
supersymmetry.  One is to simply accept a fine-tuning to maintain 
the hierarchy against radiative corrections, at a level significantly 
worse than a percent.  Quite often, the anthropic principle is brought 
in to justify this level of fine-tuning~\cite{ArkaniHamed:2004fb}. 
The second is to keep superparticles relevant to the Higgs mass 
parameter below TeV while to assume all other superparticles well 
beyond TeV~\cite{Dimopoulos:1995mi}.  The third is to assume that 
all superparticles are nearly degenerate making them somewhat hidden 
from experimental searches due to low $Q$-values in visible and 
missing energies.  The last option, however, has been discussed 
only phenomenologically~\cite{LeCompte:2011fh}, lacking theoretical 
justifications based on simple and explicit models of supersymmetry 
breaking.

In this Letter, we point out that the third possibility of a nearly 
degenerate superparticle spectrum is quite automatic when supersymmetry 
is broken by boundary conditions in compact extra dimensions, the so-called 
Scherk--Schwarz mechanism~\cite{Scherk:1978ta,Pomarol:1998sd}.  With 
the simplest extra dimension---the $S^1/{\mathbb Z}_2$ orbifold---the 
mechanism has a rather simple structure~\cite{Barbieri:2001yz}.  In 
particular, locating matter and Higgs fields in the bulk and on a brane, 
respectively, and forbidding local-parity violating bulk mass parameters 
for the matter fields, the theory has only four parameters relevant 
for the spectrum of superparticles:\ the compactification scale $1/R$, 
the 5D cutoff scale $\Lambda$ ($> 1/R$), the supersymmetry-breaking 
twist parameter $\alpha$ ($\in [0, \frac{1}{2}]$), and the supersymmetric 
Higgs mass $\mu$.

Using the common notation in the MSSM, the spectrum of superparticles 
is given at the compactification scale $\approx 1/R$ as
\begin{equation}
\begin{array}{c}
  M_{1/2} = \frac{\alpha}{R},
\quad
  m_{\tilde{Q},\tilde{U},\tilde{D},\tilde{L},\tilde{E}}^2 
  = \left( \frac{\alpha}{R} \right)^2,
\quad
  m_{H_u,H_d}^2 = 0,
\\[10pt]
  A_0 = -\frac{2\alpha}{R},
\quad
  \mu \neq 0,
\quad
  B = 0,
\end{array}
\label{eq:tree}
\end{equation}
at tree level.  While these masses receive radiative corrections from 
physics at and above $1/R$, they are under control because of the 
symmetries in higher-dimensional spacetime, and thus can naturally 
be small.  Therefore, in this limit, the theory essentially has only 
three free parameters:
\begin{equation}
  \frac{1}{R},
\qquad
  \frac{\alpha}{R},
\qquad
  \mu.
\label{eq:free-param}
\end{equation}
This rather compact set of parameters gives all the superparticle as 
well as the Higgs boson masses.

Even though Eq.~(\ref{eq:free-param}) gives two less parameters than in 
the traditional CMSSM framework, we show that it still leads to viable 
phenomenology.  In addition, it solves the flavor problem that often 
plagues models of supersymmetry breaking, because the geometry is 
universal to all scalar particles and hence respect a large flavor 
symmetry.  The problem of accommodating a large enough Higgs boson 
mass is ameliorated by the near-degeneracy between $\tilde{t}_L$ and 
$\tilde{t}_R$, and $|A_t| \approx 2 m_{\tilde{t}}$.  And the degenerate 
spectrum at tree level automatically achieves a compact spectrum that 
allows superparticles to be hidden from the current searches even 
when they are below TeV.

This Letter is organized as follows.  We first review the basics of 
supersymmetry breaking by boundary conditions in the $S^1/{\mathbb Z}_2$ 
orbifold, and present the simplest model we study.  We then present 
the low-energy spectrum of superparticles and discuss its phenomenology. 
We also provide benchmark points useful for further phenomenological 
studies of the model.

{\it Supersymmetry Breaking by Boundary Conditions.}~---~We consider 
a single compact extra dimension with the coordinate $y$ identified 
under $T: y \rightarrow y + 2\pi R$ and $P: y \rightarrow -y$.  These 
two operations satisfy the algebra $P T P = T^{-1}$ and $P^2=1$, and 
the resulting extra dimension is an interval $y \in [0, \pi R]$:\ the 
$S^1/{\mathbb Z}_2$ orbifold.

We consider a supersymmetric $SU(3)_C \times SU(2)_L \times U(1)_Y$ 
gauge theory in this spacetime, with the gauge and three generations 
of matter supermultiplets propagating in the bulk.  The boundary 
conditions for these fields are given such that the $SU(2)_R$ doublets 
in these multiplets transform as
\begin{equation}
  P = \left(
    \begin{array}{cc}
      1 & 0 \\ 0 & -1
    \end{array} \right),
\quad
  T = \left(
    \begin{array}{cc}
      \cos (2\pi\alpha) & -\sin(2\pi\alpha) \\ 
      \sin(2\pi\alpha) & \cos(2\pi\alpha)
    \end{array} \right),
\label{eq:b-c}
\end{equation}
under $P$ and $T$ (see Ref.~\cite{Barbieri:2001yz} for details).  In this 
paper we consider the case $\alpha \ll 1$.%
\footnote{The twist parameter $\alpha$ in the boundary conditions 
 is equivalent to an $F$-term vacuum expectation value of the radion 
 superfield~\cite{Marti:2001iw}, which can be generated dynamically 
 through a radion stabilization mechanics and hence can be naturally 
 small.}

The boundary conditions of Eq.~(\ref{eq:b-c}) leave only the MSSM gauge 
and matter fields below the compactification scale $1/R$.  Specifically, 
the matter supermultiplets yield three generations of quarks and leptons 
as the zero modes, while their superpartners obtain the common soft mass 
of $\alpha/R$.  (Here, we have assumed that there are no 5D bulk mass 
terms for the matter multiplets.%
\footnote{This assumption can be justified by a local parity in the bulk; 
 see, e.g.,~\cite{Barbieri:2002sw}.})
The gauge supermultiplets give massless standard model gauge fields and 
gauginos of mass $\alpha/R$.  We therefore obtain the first two expressions 
in Eq.~(\ref{eq:tree}).  (The Kaluza--Klein excitations form $N=2$ 
supermultiplets and have masses $\approx n/R$ ($n = 1,2,\cdots$), with 
supersymmetry-breaking mass splitting of order $\alpha/R$.)

The Higgs chiral superfields $H_u$ and $H_d$ are located on one of the 
branes at $y=0$.  The Yukawa couplings and $\mu$ term can then be written 
on that brane:
\begin{eqnarray}
  {\cal L}_{\rm brane} &=& \delta(y) \int\!d^2\theta \bigl( 
    y_U^{ij} Q_i U_j H_u + y_D^{ij} Q_i D_j H_d
\nonumber\\
  && {} \qquad + y_E^{ij} L_i E_j H_d + \mu H_u H_d \bigr).
\label{eq:brane-W}
\end{eqnarray}
This leads to the other four expressions in Eq.~(\ref{eq:tree}).  (Here, 
we have simply assumed the existence of the $\mu$ term on the brane.  We 
leave discussions of its origin to future work.)

Note that the degeneracy among the three generations of scalars is 
automatic because of the geometric nature of the supersymmetry breaking. 
In addition, since $\alpha$ and $\mu$ can always be taken real by phase 
redefinitions of fields associated with $R$ and Peccei-Quinn rotations, 
there is no physical phase in $M_{1/2}$, $A_0$, $\mu$, or $B$.  Therefore, 
the flavor problem as well as the $CP$ problem are automatically solved 
in this model.

The expressions in Eq.~(\ref{eq:tree}) receive corrections from 
physics above and at $1/R$.  In the 5D picture, corrections above 
$1/R$ come from brane-localized kinetic terms for the gauge and matter 
supermultiplets, and affect $M_{1/2}$, $m_{\tilde{f}}^2 \equiv 
m_{\tilde{Q},\tilde{U},\tilde{D},\tilde{L},\tilde{E}}^2$, and $A_0$. 
These terms have tree-level contributions at $\Lambda$ and radiative 
ones between $1/R$ and $\Lambda$.  From dimensional analysis, the 
size of the radiative contributions is
\begin{equation}
  \frac{\delta M_{1/2}}{M_{1/2}},\, 
    \frac{\delta m_{\tilde{f}}^2}{m_{\tilde{f}}^2},\, 
    \frac{\delta A_0}{A_0}
  \approx O\left(\frac{g^2,\, y^2}{16\pi^2} \ln(\Lambda R)\right).
\label{eq:corr-1}
\end{equation}
Moreover, it is (technically) natural to assume that the tree-level 
contributions do not exceed the radiative ones with $\ln(\Lambda R) 
\rightarrow O(1)$.  Therefore, with this assumption, the corrections 
to Eq.~(\ref{eq:tree}) from physics above $1/R$ are always negligible 
for $\Lambda R \lesssim 16\pi^2$, i.e.\ when our effective higher 
dimensional field theory is valid.  (The same can also be seen in 
the 4D picture.  In this picture, $N=2$ supersymmetry existing for 
the $n>0$ modes leads to nontrivial cancellations of the corrections 
to $M_{1/2}$, $m_{\tilde{f}}^2$ and $A_0$ from these modes.  In order 
to see the cancellations for the gauge multiplets, the effect of anomalies 
must be taken into account correctly.  The explicit demonstration 
of these nontrivial cancellations will be given elsewhere.)

The corrections from physics at $1/R$ arise from nonlocal operators in 
5D.  They affect all the supersymmetry-breaking masses, and are of order 
$1/16\pi^2$.  Here we calculate only the contributions to the Higgs mass 
parameters, which could potentially affect the analysis of electroweak 
symmetry breaking.  By choosing the renormalization scale to be $1/(2\pi R)$, 
we find these corrections are
\begin{eqnarray}
\begin{array}{l}
  \delta m_{H_u}^2 
  = \left( -\frac{33 y_t^2}{8\pi^2} 
    + \frac{9 (g_2^2 + g_1^2/5)}{16\pi^2} \right) 
    \left( \frac{\alpha}{R} \right)^2,
\\[5pt]
  \delta m_{H_d}^2 
  = \frac{9 (g_2^2 + g_1^2/5)}{16\pi^2} 
    \left( \frac{\alpha}{R} \right)^2,
\\[5pt]
  \delta B 
  = \left( \frac{9 y_t^2}{8\pi^2} 
    - \frac{3 (g_2^2 + g_1^2/5)}{8\pi^2} \right) 
    \frac{\alpha}{R},
\end{array}
\label{eq:corr-2}
\end{eqnarray}
where we have included only the contributions from the top-Yukawa coupling, 
$y_t$, and $SU(2)_L$ and $U(1)_Y$ gauge couplings, $g_2$ and $g_1$ (in 
the $SU(5)$ normalization).

In summary, the low-energy superparticle masses are obtained by evolving 
down Eqs.~(\ref{eq:tree},~\ref{eq:corr-2}) defined at the renormalization 
scale
\begin{equation}
  \mu_{\rm RG} = \frac{1}{2\pi R},
\label{eq:RG-scale}
\end{equation}
using the MSSM renormalization group equations.  Incidentally, 
the gravitino mass is $m_{3/2} = \alpha/R$, generated by the 
supersymmetry-breaking twist in the fifth dimension.

\begin{figure}[t]
\begin{center}
  \includegraphics[clip,width=.45\textwidth]{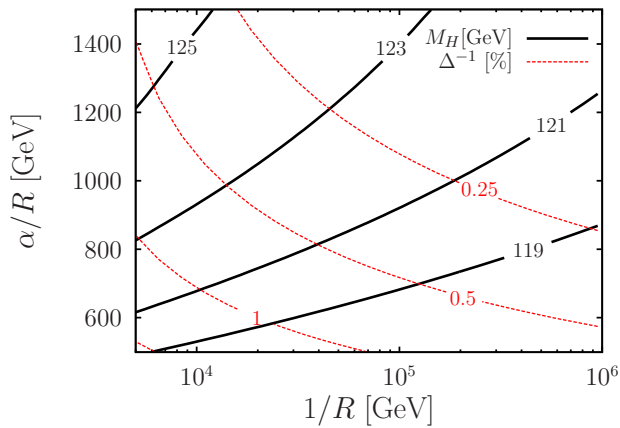}
\end{center}
\caption{The lightest Higgs boson mass $M_H$ (in GeV) and the fine-tune 
 parameter $\Delta^{-1}$.  Note that there is an approximately $3~{\rm GeV}$ 
 systematic error in theoretical computation of $M_H$.}
\label{fig:higgs}
\end{figure}

{\it Superparticle Spectrum.}~---~Following the procedure 
described above, we calculate the MSSM mass spectrum using 
{\tt SOFTSUSY~3.3.1}~\cite{Allanach:2001kg} and the lightest Higgs 
boson mass using {\tt FeynHiggs~2.8.6}~\cite{Heinemeyer:1998yj}.  In 
Fig.~\ref{fig:higgs}, we plot the contours of the mass of the lightest 
Higgs boson, $M_H$, and the fine-tune parameter, defined by $\Delta^{-1} 
= {\rm min}_x |\partial \ln m_Z^2 / \partial \ln x|^{-1}$ with $x = 
\alpha, \mu, 1/R, y_t, g_3, \cdots$, in the $1/R$-$\alpha/R$ plane.
(The fine-tuning parameter is determined mostly by $x = \mu$.) 
In the calculation, we have used the top-quark mass of $m_t = 
173.2~{\rm GeV}$~\cite{Lancaster:2011wr}.  Varying it by $1\sigma$, 
$\varDelta m_t = \pm 0.9~{\rm GeV}$, affects the Higgs boson mass 
by $\varDelta M_H \approx \pm 1~{\rm GeV}$.  Also, theoretical 
errors in $M_H$ are expected to be about $|\varDelta M_H| \approx 
2~\mbox{--}~3~{\rm GeV}$~\cite{Allanach:2004rh}, so that the regions 
with $M_H \gtrsim 121~\mbox{--}~123~{\rm GeV}$ in the plot are not 
necessarily incompatible with the $125~{\rm GeV}$ Higgs boson hinted 
at the LHC~\cite{ATLAS:2012ae}.  Indeed, using the recently-released 
program {\tt H3m}~\cite{Kant:2010tf}, which includes a partial three-loop 
effect, we find that the corrections to $M_H$ from higher order effects 
are positive and of order a few GeV in most of the parameter region in 
the plot.

\begin{figure}[t]
\begin{center}
  \includegraphics[clip,width=.45\textwidth]{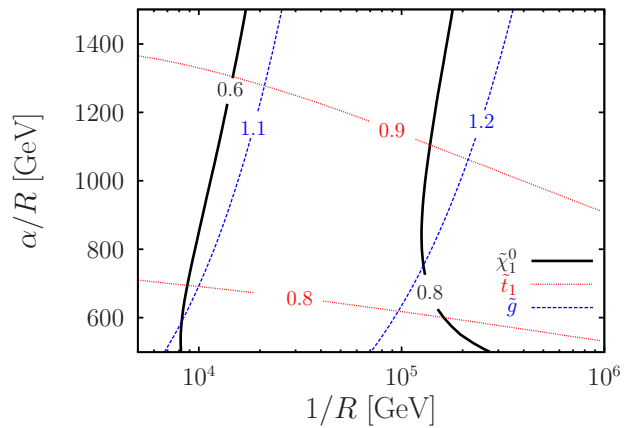}
\end{center}
\caption{Masses of the lightest neutralino $\tilde{\chi}^0_1$, the lighter 
 top squark $\tilde{t}_1$, and the gluino $\tilde{g}$ normalized to 
 $\alpha/R$.}
\label{fig:mass}
\end{figure}

In Fig.~\ref{fig:mass}, the masses of selected superparticles (the 
lightest neutralino $\tilde{\chi}^0_1$, the lighter top squark 
$\tilde{t}_1$, and the gluino $\tilde{g}$) are shown.
The masses of the first and second generation squarks are almost the 
same as the gluino mass.  The masses of the electroweak superparticles 
are close to $\alpha/R$, except for the lightest two neutralinos 
$\tilde{\chi}^0_{1,2}$ and the lighter chargino $\tilde{\chi}^+_1$, 
which are Higgsino-like (and thus close in mass) in most of the parameter 
space.  We find that the masses of the superparticles are degenerate 
at a $10\%$ level, except possibly for the Higgsinos which can be 
significantly lighter (up to a factor of $\approx 2$).

{\it Experimental Limits.}~---~As we have seen, the model naturally 
predicts a degenerate mass spectrum for superparticles.  This has 
strong implications on supersymmetry searches at the LHC.  Because 
of the mass degeneracy, production of high $p_{\rm T}$ jets and large 
missing energy is suppressed.  Therefore, typical searches, based on 
high $p_{\rm T}$ jets and large missing energy, are less effective for 
the present model.

\begin{figure}[t]
\begin{center}
  \includegraphics[clip,width=.45\textwidth]{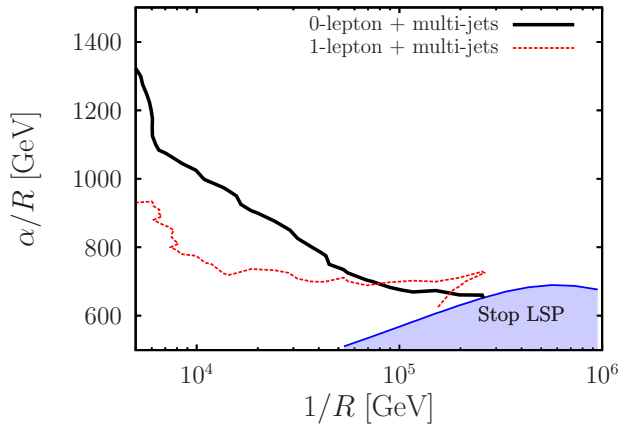}
\end{center}
\caption{The current LHC constraint on the model.}
\label{fig:LHC}
\end{figure}

To estimate the number of supersymmetric events, we have used 
{\tt ISAJET~7.72}~\cite{Paige:2003mg} for the decay table of superparticles, 
{\tt Herwig~6.520}~\cite{Corcella:2000bw} for the generation of 
supersymmetric events, {\tt AcerDET~1.0}~\cite{RichterWas:2002ch} 
for the detector simulation, and {\tt NLL-fast}~\cite{NLLFAST} for 
estimation of the production cross section including next-to-leading
order QCD corrections and the resummation at next-to-
leading-logarithmic accuracy.  To constrain the 
parameter space, we compare the obtained event numbers with the results 
of ATLAS searches for multi-jets plus large missing energy with and 
without a lepton at $L = 4.7~{\rm fb}^{-1}$ at $\sqrt{s} = 7~{\rm 
TeV}$~\cite{ATLAS1,ATLAS2}.  In Fig.~\ref{fig:LHC}, we show the resulting 
LHC constraint on the model.
Other searches such as those for $b$-jets and/or multi-leptons are less 
effective.  We find that for $1/R \gtrsim 10^5~{\rm GeV}$, the case 
that $m_{\tilde{g}} \simeq m_{\tilde{q}} \lesssim 1~{\rm TeV}$ is 
still allowed.  This constraint is significantly weaker than that 
on the CMSSM, which excludes $m_{\tilde{g}} \lesssim 1.4~{\rm TeV}$ 
for $m_{\tilde{g}} \simeq m_{\tilde{q}}$~\cite{ATLAS1}.  (We have 
checked that our naive method of estimating the LHC constraints adopted 
here reproduces this bound for the CMSSM spectra.)

We note that since $B$ is not a free parameter in the present model, 
$\tan\beta$ is determined by the electroweak symmetry breaking condition. 
We typically find $\tan\beta \sim 4~\mbox{--}~10$.  This allows for the 
model to avoid the constraint from $b \rightarrow s\gamma$, despite the 
large $A$ terms.

The contribution of the Kaluza--Klein states to the electroweak precision 
parameters bounds $1/R \gtrsim \mbox{a few}~{\rm TeV}$~\cite{Delgado:2001si}. 
Since we consider the region $1/R \gtrsim 10~{\rm TeV}$ in this paper, 
however, the model is not constrained by the electroweak precision data.

\begin{figure}[t]
\begin{center}
  \includegraphics[clip,width=.45\textwidth]{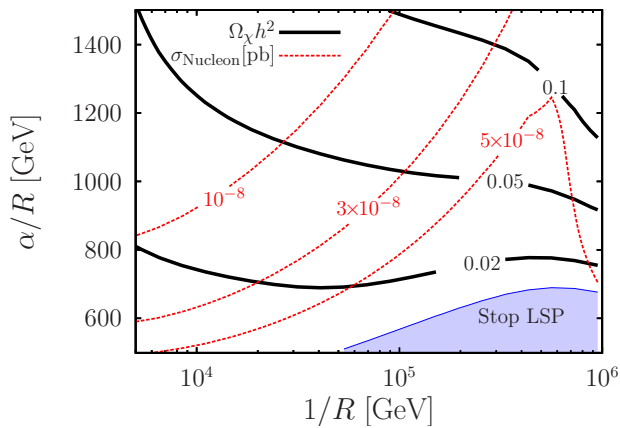}
\end{center}
\caption{The thermal relic abundance, $\Omega_\chi h^2$, and the 
 spin-independent cross section with a nucleon, $\sigma_{\rm Nucleon}$, 
 of $\tilde{\chi}^0_1$.  The solid (black) lines are the contours of 
 $\Omega_\chi h^2$, while the dotted (red) lines are those of 
 $\sigma_{\rm Nucleon}$.}
\label{fig:DM}
\end{figure}

{\it Dark Matter.}~---~In the present model, the dark matter candidate is 
the lightest neutralino $\tilde{\chi}^0_1$, whose dominant component is 
the Higgsino.  In Fig.~\ref{fig:DM}, we show the thermal relic abundance, 
$\Omega_\chi h^2$, and the spin-independent cross section with a nucleon, 
$\sigma_{\rm Nucleon}$, of $\tilde{\chi}^0_1$, assuming $R$-parity 
conservation.
To estimate these, we have used {\tt micrOMEGAs~2.4}~\cite{Belanger:2010gh}. 
For the strange quark form factor we have adopted $f_s = 0.02$, suggested 
by lattice calculations~\cite{Ohki:2008ff}, instead of the default value 
of {\tt micrOMEGAs} ($f_s = 0.26$).  As seen in Fig.~\ref{fig:DM}, the 
thermal relic abundance of $\tilde{\chi}^0_1$ is much smaller than the 
observed dark matter density $\Omega_{\rm DM} h^2 \simeq 0.1$, unless 
$\tilde{\chi}^0_1$ is rather heavy $\sim {\rm TeV}$ (in the upper-right 
corner of the plot).  Therefore, in most parameter regions, 
$\tilde{\chi}^0_1$ cannot be the dominant component of dark matter 
if only the thermal relic abundance is assumed.  It must be produced 
nonthermally to saturate $\Omega_{\rm DM} h^2$, or some other 
particle(s), e.g.\ the axion/axino, must make up the rest.

\begin{figure}[t]
\begin{center}
  \includegraphics[clip,width=.45\textwidth]{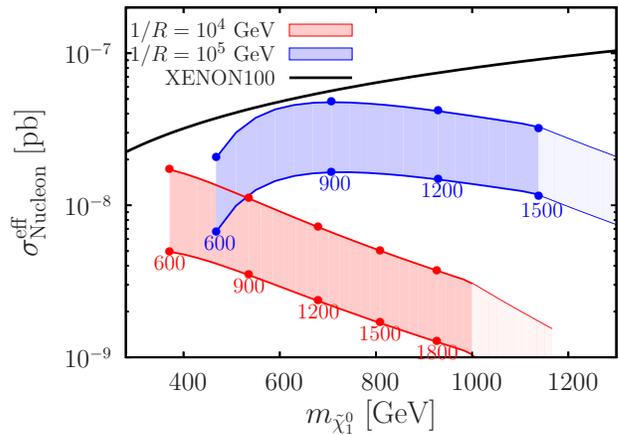}
\end{center}
\caption{Effective dark matter-nucleon cross section for $1/R = 10^4~{\rm 
 GeV}$ (lower, red shaded) and $10^5~{\rm GeV}$ (upper, blue shaded). 
 In each region, the upper and lower boarders correspond to $f_s=0.26$ 
 and $0.02$, respectively, and the dots represent the corresponding 
 values of $\alpha/R$. (The very light shaded regions are those in which 
 the thermal abundance exceeds $\Omega_{\rm DM}$.)  The solid (black) 
 line shows the current upper bound from XENON100.}
\label{fig:eff}
\end{figure}

It is natural, however, to expect that at least the thermal abundance 
of $\tilde{\chi}^0_1$ remains as a (sub-)component of dark matter.  In 
this case, direct and indirect signatures of the relic neutralino are 
expected.  To discuss the direct-detection signal, is is useful to define
\begin{equation}
  \sigma_{\rm Nucleon}^{\rm eff} \equiv \sigma_{\rm Nucleon} 
    \frac{{\rm min}\{\Omega_\chi, \Omega_{\rm DM}\}}{\Omega_{\rm DM}},
\label{eq:sigma_eff}
\end{equation}
which is the quantity to be compared with the dark matter-nucleon cross 
section in the usual direct-detection exclusion plots (which assume 
$\Omega_\chi = \Omega_{\rm DM}$).  In Fig.~\ref{fig:eff}, we plot 
$\sigma_{\rm Nucleon}^{\rm eff}$ as a function of $m_{\tilde{\chi}^0_1}$ 
for $1/R = 10^4~{\rm GeV}$ and $10^5~{\rm GeV}$.
To represent the uncertainty from the nucleon matrix element, we show both 
the $f_s=0.02$ and $0.26$ cases.  We also present the current upper bound 
on $\sigma_{\rm Nucleon}^{\rm eff}$ from XENON100~\cite{Aprile:2011hi}. 
We find that improving the bound by one or two orders of magnitude will 
cover a significant portion of the parameter space of the model.

{\it Final Remarks.}~---~In this Letter, we pointed out that supersymmetry 
broken by boundary conditions in extra dimensions leads naturally to a nearly 
degenerate superparticle spectrum, ameliorating the limits from experimental 
searches.  We presented the simplest such model in the $S^1/{\mathbb Z}_2$ 
orbifold, and showed that it leads to viable phenomenology despite the fact 
that it essentially has two less free parameters than the CMSSM: $1/R$, 
$\alpha/R$, and $\mu$.  In Table~\ref{tab:masses} we give two representative 
points in the parameter space, which can serve benchmark points for 
further phenomenological studies.
\begin{table}[t]
\caption{Phenomenologically viable mass spectrum of the benchmark points 
 (in GeV).  Point1: $1/R = 10^4~{\rm GeV}$, $\alpha/R = 1400~{\rm GeV}$ 
 and Point2: $1/R = 10^5~{\rm GeV}$, $\alpha/R = 800~{\rm GeV}$.}
\begin{center}
\begin{tabular}{lcc|lcc}
\hline
Particle &  Point1 & Point2 & Particle & Point1 & Point2 
\\ \hline
$\tilde{g}$          & $1494$ & $949$ &                 --   &     -- &    -- \\
$\tilde{u}_L$        & $1467$ & $939$ & $\tilde{u}_R$        & $1459$ & $925$ \\
$\tilde{d}_L$        & $1469$ & $942$ & $\tilde{d}_R$        & $1458$ & $924$ \\
$\tilde{b}_2$        & $1460$ & $924$ & $\tilde{b}_1$        & $1430$ & $875$ \\
$\tilde{t}_2$        & $1557$ & $988$ & $\tilde{t}_1$        & $1267$ & $681$ \\
$\tilde{\nu}$        & $1411$ & $822$ & $\tilde{\nu}_\tau$   & $1410$ & $822$ \\
$\tilde{e}_L$        & $1413$ & $826$ & $\tilde{e}_R$        & $1406$ & $812$ \\
$\tilde{\tau}_2$     & $1417$ & $823$ & $\tilde{\tau}_1$     & $1402$ & $809$ \\
$\tilde{\chi}^0_1$   &  $767$ & $630$ & $\tilde{\chi}^0_2$   &  $777$ & $671$ \\
$\tilde{\chi}^0_3$   & $1384$ & $755$ & $\tilde{\chi}^0_4$   & $1410$ & $821$ \\
$\tilde{\chi}^\pm_1$ &  $771$ & $642$ & $\tilde{\chi}^\pm_2$ & $1409$ & $817$ \\
$h^0$                &  $125$ & $120$ & $H^0$                &  $819$ & $718$ \\
$A^0$                &  $819$ & $717$ & $H^\pm$              &  $822$ & $722$ \\
\hline
\end{tabular}
\end{center}
\label{tab:masses}
\end{table}

The theory presented here can be extended in several different ways. 
An interesting one is to introduce a singlet field $S$ together with 
superpotential interactions on the $y=0$ brane: $\lambda S H_u H_d + f(S)$, 
where $f(S)$ is a polynomial of $S$ with the simplest possibility being 
$f(S) = -\kappa S^3/3$.  This allows for an extra contribution to the 
Higgs boson mass from $\lambda$, and can make the lightest neutralino 
(which would now contain a singlino component as well) saturate the 
observed dark matter abundance without resorting to nonthermal production. 
Detailed studies of this possibility will be presented elsewhere.

\acknowledgments
This work was supported in part by the DOE under contract DE-AC02-05CH11231. 
The work of HM was also supported by the NSF under grant PHY-1002399, the 
JSPS grant (C) 23540289, the FIRST program ``Subaru Measurements of Images 
and Redshifts (SuMIRe),'' CSTP, and by WPI, MEXT, Japan.  The work of YN 
was supported in part by the NSF under grant PHY-0855653.  The work of 
KT was supported in part by the Grant-in-Aid for JSPS Fellows, and KT 
acknowledges the financial support of Friends of Todai, Inc.\ to conduct 
research in Berkeley.

\end{document}